# TunableTi:Al$_2$O$_3$ oscillator optimized for high-repetition-rate and short pulses

**Jürgen Sawinski, Delphine Débarre, Winfried Denk**

Max-Planck-Institute for Medical Research, Dept. Biomedical Optics,

Jahnstr. 29, 69120 Heidelberg, Germany

A laser was designed and constructed with the goal of producing ultra-short pulses at a high repetition rate as needed for certain applications of multi-photon microscopy. With pure prism-based dispersion compensation repetition rates of up to 270MHz were achieved. The laser operates with hard- and soft-aperturing at the third (diverging output) and the first (parallel output) stability limits, respectively. At the third stability limit we found a pulse width of 27fs (FWHM) at 800nm central wavelength. At the first stability limit pulse widths of 23-40fs with tunability from 780nm to 920nm were reached.

*PCAS: 42.60.By Design of specific laser systems | 42.62.Be Biological and medical applications | 87.64.Vv Multiphoton microscopy*

# 1. Introduction

Efficient multi-photon laser-scanning microscopy (MPLSM) requires pulsed excitation because the efficiency of non-linear optical processes depends on the averaged n-th power of the instantaneous laser intensity. For 2-photon excitation this means that at constant average power the average excitation rate increases with the reciprocal duty cycle ($I \propto P_{mean}^2 / f\tau$, with the average fluorescence intensity $I$, the repetition rate $f$ and the pulse width $\tau$). While the duty cycle can, of course, be reduced by lowering the pulse-repetition rate [1] a high repetition rate is often preferred [2]. The use of femto-second pulses is, therefore, standard in multi-photon microscopy. In fact, commercially available mode-locked Ti:Sapphire lasers are the excitation-light source used for the vast majority of multi-photon microscopes in operation. The repetition rates of such lasers (around 100MHz) provide a sufficient number of pulses per pixel to avoid the need for synchronization between pulses and acquisition clock. With an easily achievable pulse width of around 100fs the inverse duty cycle (and hence enhancement of 2-photon absorption) is about $10^5$.

While for most applications such pulse parameters are appropriate, sometimes higher repetition rates are desirable or even necessary: (1) when saturation due to ground-state depletion limits the achievable fluorescence intensity [3], (2) when super-linear bleaching [4] or photodamage need to be minimized but a certain fluorescence intensity is needed, such as in fluorescence correlation spectroscopy [5], or (3) when excitation through optical fibers is necessary, where lowering the pulse energy by increasing the repetition rate allows a reduction of nonlinear pulse broadening [6]. To make up for the reduction of the excitation efficiency due to the increased repetition rate shorter pulses can be used. Somewhat surprisingly, shorter pulses may, furthermore, reduce the severity of non-linear effects in fibers as the length of fiber over which high peak powers persist is shorter due to the larger spectral width [7]. While very high repetition rates (in the GHz regime) can be achieved by using chirped-mirror dispersion compensation [8, 9] this entails a loss of broad tunability, which is needed for multi-photon optical imaging to allow the excitation of different fluorophores. We, therefore, custom designed a laser optimized for high repetition rates and short pulses with entirely prism-based dispersion compensation (compare [10]).

## 2. Design

A major factor limiting the repetition rate of mode-locked lasers is the optical path-length required for prism-based group velocity dispersion (GVD) compensation [11]. This is particularly true for very short pulse durations where fused-silica prisms need to be used since prisms made from high-dispersion materials generate too much third-order dispersion (TOD) [12]. To reduce the required inter-prism distance we used two pairs of prisms [13] resulting in a reduction of the required apex distance by roughly a factor of four [14]. When using a 5mm long Ti:Sapphire crystal the required inter-prism distance for GVD compensation is ≈200mm. Mechanical constrains add about 100mm leading to a total length of 300mm of the $L_2$ arm of the cavity (for a drawing of the cavity layout, see Fig. 1).

The $L_1$ arm also requires a minimum length if beam propagation is not parallel, as is the case for mode-locking using a hard-aperture, which needs to be placed near the end of $L_1$ at the output coupler and required operation near the third stability limit of the cavity. The theoretical lower limit for $L_1$ is then twice the focal length of $M_2$, in our case $L_1$=100mm. In practice we could reliably maintain mode-locking only down to about $L_1$=200mm (see below). An alternative is to operate the laser at the first stability limit, then, of course, without a hard aperture (Fig. 1), and no limit on L1. We designed and operated cavity variants for both 3rd and 1st stability-limit operation.

## 2.1. Dispersion compensation

The double-prism sequence we used for GVD compensation is often referred to as the Proctor and Wise arrangement [13]. To calculate the needed inter-prism distance we initially used analytical expressions derived by Sherriff [15] for prism sequences that are more complicated than the original Gordon and Fork [16] arrangement. Assuming a 5mm long Ti:Al$_2$O$_3$ crystal (where only material dispersion is considered, compare [17]) and a beam that is passing 4mm of glass at each prism at the central wavelength we find an inter-prism distance (IPD, apex P$_2$ to apex P$_3$) of 187mm needed to balance the cavity GVD at 850nm, the design wavelength of the prism arrangement. The Sherriff formulas are, however, only valid at minimum deviation [14] (see also [18]) and adjusting the prism angles to keep them at minimum deviation while tuning the laser wavelength is impractical. We, therefore, used a numerical ray-tracing technique instead to calculate the dispersion (for an analytical solution for truly arbitrary prism sequences, see [19]).

As a side-effect the ray-tracing allowed us to follow rays at the beam edges and check for clipping: for the design of the laser we specified that a beam with 4mm diameter and found that a bandwidth of ±30nm should just remain inside the prisms apexes. In this case the required IPDs become 239, 228, 209, 186 and 159mm for 750, 800, 850, 900, and 950nm, respectively. The corresponding GVD and TOD curves are shown in Fig. 2. The residual TOD, which limits the shortest achievable pulse width, is below 1700fs$^3$ in the range from 750 to 980nm, which should allow pulses below 30fs [20].

*2.2. Cavity optimization*

The cavity was modeled using the ABCD matrix ansatz [21] in a two-step approach: first we calculated the stability regions as a function of the folding mirror distance (z), with $L_1$ and $L_2$ as parameters without, however, taking optical non-linearity into account. Since the crystal faces are traversed and the curved mirrors reflect at non-normal angles we have to treat the tangential and sagittal planes separately [22]. For the first stability limit both folding angles have to be identical if the beam inside the crystal is to be stigmatic ($\varphi_3=\varphi_2=17.1°$) and the folding mirror distance is then $z_1=108$mm. For the third stability limit the folding angle ($\varphi_3$) for the arm with the parallel beam ($L_2$) is unchanged if the beam inside the crystal is to remain stigmatic and we can calculate $\varphi_2$ using again the condition that the tangential and sagittal stability limits have to occur at the same z ($z_3=125$mm). We find $\varphi_2=13.2°$ for an arm length of $L_1=200$mm (unlike for the first stability limit, $\varphi_2$ does depend on the arm length, with values ranging from 8.5° to 14.2° for L1=100mm to 300mm).

In a second step we introduced Kerr lensing, represented by an additional ABCD matrix. We then calculated for a given z that position (x) of the crystal for which the maximal intensity dependence of the beam diameter at the output coupler occurs [23]. We found x = 50.45mm both at the first and at the third stability limit; we also found (Fig. 3) that the mode variation is dominant in the tangential plane, which is essential for a vertically orientated slit aperture.

## 3. Cavity construction

The cavity focusing mirrors have a radius of curvature of 100mm; the output coupler has a transmission coefficient of 12% (all cavity mirrors used were from a broad-band "X-Wave" mirror set for a Mira 900 Laser, Coherent GmbH, Dieburg, Germany). The prisms are fused silica with an apex angle of α=68.7° (IB-10.5-68.7-SS, CVI Laser Corporation, New Mexico,

USA). The gain medium was a short (5mm) highly doped ($\alpha=3.4\text{cm}^{-1}$) Ti:sapphire crystal cut at Brewster's angle (MolTech GmbH, Berlin, Germany) pumped with up to 10.5W of the 532nm light from a diode-pumped frequency doubled Nd:YVO$_4$ laser (Verdi V-10, Coherent Inc.). The focusing mirror for the pump beam was taken from a Tsunami laser (Spectra Physics).

## *3.1. Alignment*

We first optimized the cavity for CW operation without the prism sequence. The arm lengths $L_1$ and $L_2$ were set to be nearly equal, at about 300mm, with $L_1$ slightly smaller than $L_2$ to ensure a small spot on the OC at the third stability limit. For operation at the third or first stability limit we started out with z set slightly larger than the theoretical values of $z_3=125$mm and $z_1=108$mm, respectively. The end mirror tilts, crystal position, and pump focus were then adjusted for maximum output power.

For third stability-limit operation we then began to reduce $L_1$ while keeping the CW power maximized by making small adjustments to the end mirrors and to x, z, and $\varphi_2$. While the calculated values for the optimal folding angles are $\varphi_2=13.2°$ and $\varphi_3=17.1°$ for a $L_1=200$mm we found that, in practice, the folding angles are pump power dependent. Better results were obtained by decreasing the folding angles while increasing power, presumably due to astigmatic thermal lensing. For pump powers between 6 and 8W good results were obtained with $\varphi_2=12°$, and $\varphi_3=13°$. These values are not optimal, especially when $L_1$ should be shorter than 200mm, but due to mechanical constraints $\varphi_2$ could not be reduced below 12°. Residual astigmatism can strongly affect mode-locking because Kerr lensing is dominant in the tangential plane (see Fig. 3). This means that to ensure effectiveness of the slit we have to keep the tangential mode variation on the output coupler at least as large as the sagittal mode variation. In practice, this can be tested by observing the size and ellipticity of the output beam as z is decreased towards the third stability limit: with proper astigmatism compensation the beam stays round as its diameter increases with decreasing z.

At the first stability limit the folding angles ($\varphi_2=\varphi_3$) were varied between 13º and 17º. For a pump power of 10W an angle of 15º appeared to yield the best results for mode-locking operation. As we have parallel ray propagation in the $L_1$ arm, $L_1$ can be made quite small, limited only by mechanical constraints, and was initially set to 200mm and later reduced to 95mm. As expected from the calculations we did not observe any relation between $L_1$ and mode-locking ability in our experiments.

## *3.2. Prism sequence alignment*

First we inserted only the prism pair $P_1/P_2$ into the cavity. The end-mirror horizontal tilt was initially set so that the laser operated at the intended wavelength. The angular orientation of prism pair $P_1/P_2$ was then adjusted to yield a minimal operating wavelength. These steps were repeated until the minimal wavelength and intended wavelength were equal. This adjustment corresponds to finding the minimum-deflection orientation for $P_1/P_2$. The second prism pair ($P_3/P_4$) was then inserted and its angular orientation was adjusted until tilting the end mirror horizontally no longer produced a shift in the operating wavelength.

## *3.3. Mode locking*

Before attempting to start mode-locking at the third stability limit we had to readjust the folding-mirror distance (z) and the crystal position (x), which were still set for maximum power in CW-mode (z=118±2mm, x=51±1mm). Empirically we found that the laser started to mode-lock most easily when (at a pump power P=6W) z and x were increased by 0.3mm and 0.1mm, respectively. To start mode-locking we first closed the slit in front of the output coupler until the output power was reduced by 25%. The vane $S_2$, which controls the wavelength and which is located in front of $M_4$, was adjusted so that wavelengths above 805nm were blocked. Mode-locking could now easily be started by briefly and quickly pushing $P_{3/4}$ away from the beam (less glass, direction denoted by the arrow in Fig. 1).

To start model-locking at the first stability limit we found that, like for the third stability limit, the folding mirror distance z has to be increased with increasing pump power. Empirically we found that at a pump power of 10W mode-locking could be achieved by increasing z by 0.5 to 1.0mm from the maximum CW-power position (z=105±2mm). The position of the crystal was optimized by constantly moving the prism pair $P_3/P_4$ in and out of the beam and set to a position where mode-locking starts with the smallest movement.

In practice, we observed that the initial alignment of the cavity is more critical at the first than at the third stability limit (compare [24]) with slight misalignment causing mainly instability of the mode-locking process (breakdown to CW within seconds) and CW breakthrough. We, therefore, further optimized crystal position and end-mirror tilt while the laser was mode-locking. Alternatively it was possible to reduce the sharp frequency components by using the $S_2$ vane (suppression of CW parts is, however, not always possible, compare Fig. 5). Interestingly, we found that the cavity is less susceptible to dust at the first stability limit (with roughly twice as long intervals between necessary mirror cleanings).

## 4. Results

Near the third stability limit, running CW at a wavelength of 810nm the laser had conversions efficiencies of up to 15% and 17% at a pump powers of $P_p$ =6W and $P_p$=10W, respectively. The lasing-threshold was at $P_p$=2.1W.

In mode-locked operation at a center wavelength of 800nm the conversion efficiencies were 10% and 12.5% at $P_p$=6W and $P_p$=8W, respectively. All spectra and auto-correlations were recorded with GVD compensation (using two SF10 prisms) after the output coupler (Fig. 4). Chi-squared fits assuming a Gaussian pulse profile (carried out with ROOT [25]) gave a central wavelength of 797.8±0.2nm (376.1±0.1THz, mean ±SEM), a spectral width of 41.4±0.1nm (19.5±0.1THz) and a pulse width of 26.5±0.1fs. The time-bandwidth product calculated from these values is 0.52, slightly larger than the transform limit of 0.44. The pulse repetition rate was ≈240 MHz.

For an inter-prism distance of 185mm the tuning range for stable mode-locking was 750nm to 805nm at the third stability limit. Beyond 805nm the spectrum became distorted and mode-locking broke down completely beyond 810nm. While we have not investigated this in detail because we switched to the 1st stability (where external parallelization of the diverging beam is not necessary which reduces the complexity of the adjustment for fiber coupling) reducing the IPD would probably allow longer-wavelength mode-locked operation.

At the first stability limit the conversion efficiencies were similar, reaching a peak of 19% at $P_p$=10W at 810nm but the lasing-threshold was slightly higher with $P_p$=2.2W. We achieved stable mode-locked operation of the laser, at a pump power of $P_p$=10W, over a wider spectral range than when operating at the 3rd stability limit. The center wavelength mainly depended on how far the prism pair $P_3$/$P_4$ was inserted. This effect appears not to be due to wavelength dependent clipping since it is not possible to shift the central frequency by the use of the wavelength controlling vane $S_2$ without disturbing mode-locking. Instead the operating wavelength may depend on where the shortened pulses and hence the best Kerr lensing occur. The spectra typically deviate strongly from the Gaussian shape (Fig. 5). The bandwidth typically is 40-50nm between 780nm and 910nm. At 870nm a bandwidth of up to 100nm could be reached (note that this coincides with the lowest calculated TOD, compare Fig. 2). The repetition rates were 270MHz. The pulse-widths (measured with GVD compensation by a pair of fused silica prisms) were typically between 23-40fs (calculated rms width by comparison of the Fourier-transformed spectrum and the measured auto-correlation trace). We see, however, that the measured pulse width is typically twice (up to four times for the broad spectrum at 870nm central wavelength) that of the transform-limited width. The best conversion efficiencies varied from 12% to 7% for wavelengths between 800 and 920nm.

## 5. Conclusions

We have demonstrated stable operation of a high-repetition rate Ti:sapphire oscillator with prism-based dispersion compensation in different cavity configurations. We could reach sub-30fs pulses for hard-aperture (third stability limit) as well as soft-aperture (first stability limit) supported mode-locking with repetition rate of up to 270MHz. In comparison to a typical commercial lasers (80MHz, 100fs) our laser thus achieves a 3-fold higher repetition rate and a 3-fold shorter pulse length. Important for two-photon microscopy is that stable operation near 930 nm is possible where many xanthene dyes [26] as well as many fluorescent proteins can be efficiently excited [27].

Applications envisioned for this laser include situations where pulse energies have to be minimized, for example to reduce superlinear photobleaching [28] [5] [29] [30] [4] or for multi-photon fiber microscopy [31-33]. Preliminary experiments (Sawinski and Denk, unpublished data) show that we can reach pulse lengths around 100fs for an average output power of up to 300mW at 900nm center wavelength when passing the output from this laser with appropriate pre-chirping through 1m of standard single-mode fiber .

## Acknowledgements

We thank Paul French for discussions and Mazahir T. Hasan, Axel Nimmerjahn, Bernd Kuhn, and Patrick Theer for comments on the manuscript.

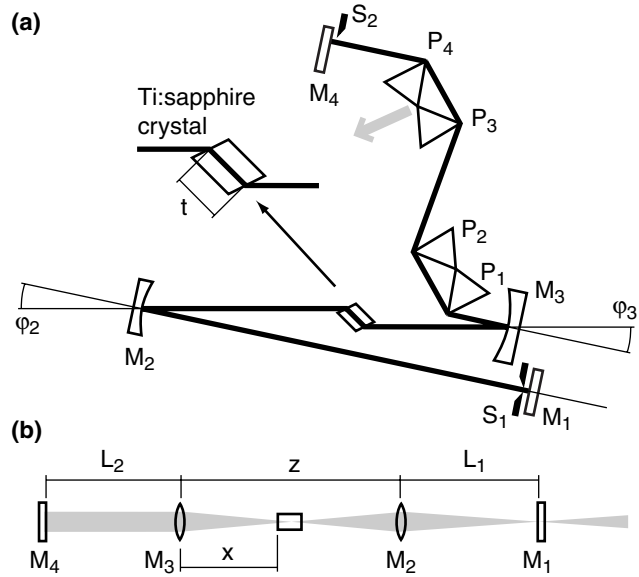

Figure 1: (a) Schematic drawing of the laser cavity: output coupler ($M_1$), focusing mirrors ($M_2$ and $M_3$), Ti:sapphire crystal, dispersion compensation consisting of four prisms ($P_1$ to $P_4$), end mirror ($M_4$) and hard-aperture slit ($S_1$) and wavelength controlling vane ($S_2$). The inter-prism distance is measured from apex to apex ($P_2$ to $P_3$). The pump beam (not shown) is focused into the crystal through $M_3$. (b) Unfolded beam path at the first and third stability limit with the focusing mirrors represented by lenses. (See text for further details.)

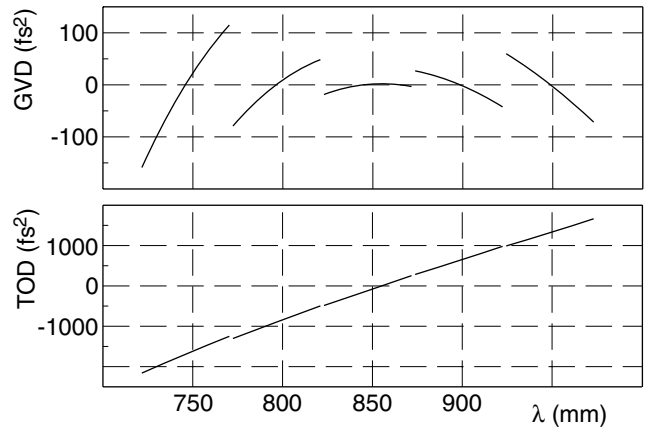

Figure 2: Calculated intra-cavity group-delay (GVD) and third-order dispersion (TOD). At the central wavelength of each partial trace the inter-prism distance was adjusted for zero GVD (to values of 239, 228, 209, 186 and 159mm, respectively).

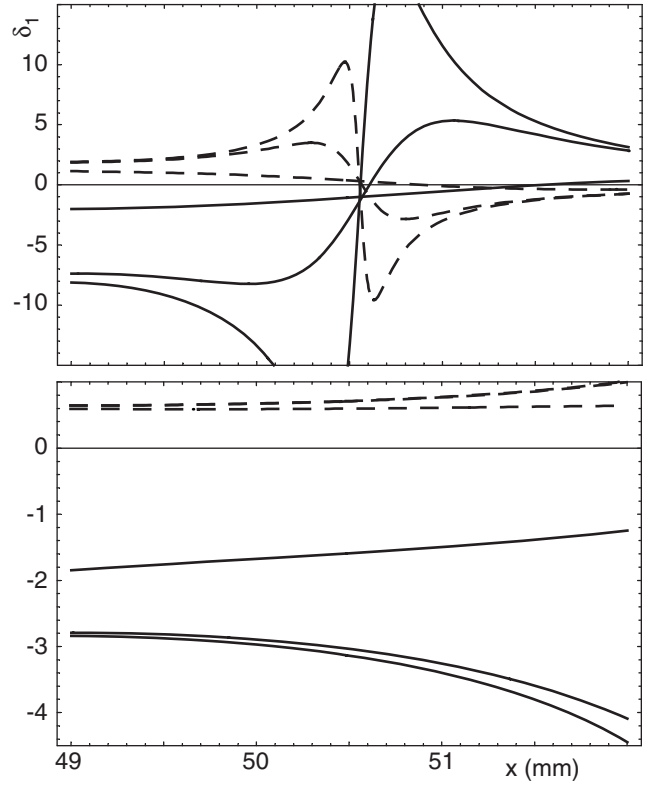

Figure 3: First order Kerr-lens induced differental size variation ($\delta_1$) of the mode on the output coupler in the tangential (*top*) and sagittal (*bottom*) plane plotted as a function on the crystal position x near the first (*dashed line*) and the third stability limit (*solid line*). Shown are data for three different offsets of z: 0.3, 0.01 and 0.001mm (increasing $\delta_1$ with vanishing z) from the respective stability limit. Note difference in y-scales.

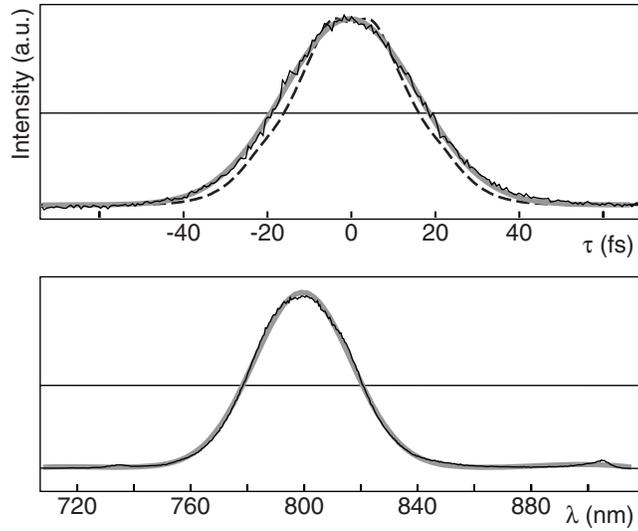

Figure 4: Auto-correlation (*top*) and spectrum (*bottom*) of laser output at the third stability limit, Gaussian fit (*gray*); Fourier transform of the spectrum (*dashed*), which corresponds to the autocorrelation expected for a transform limited pulse.

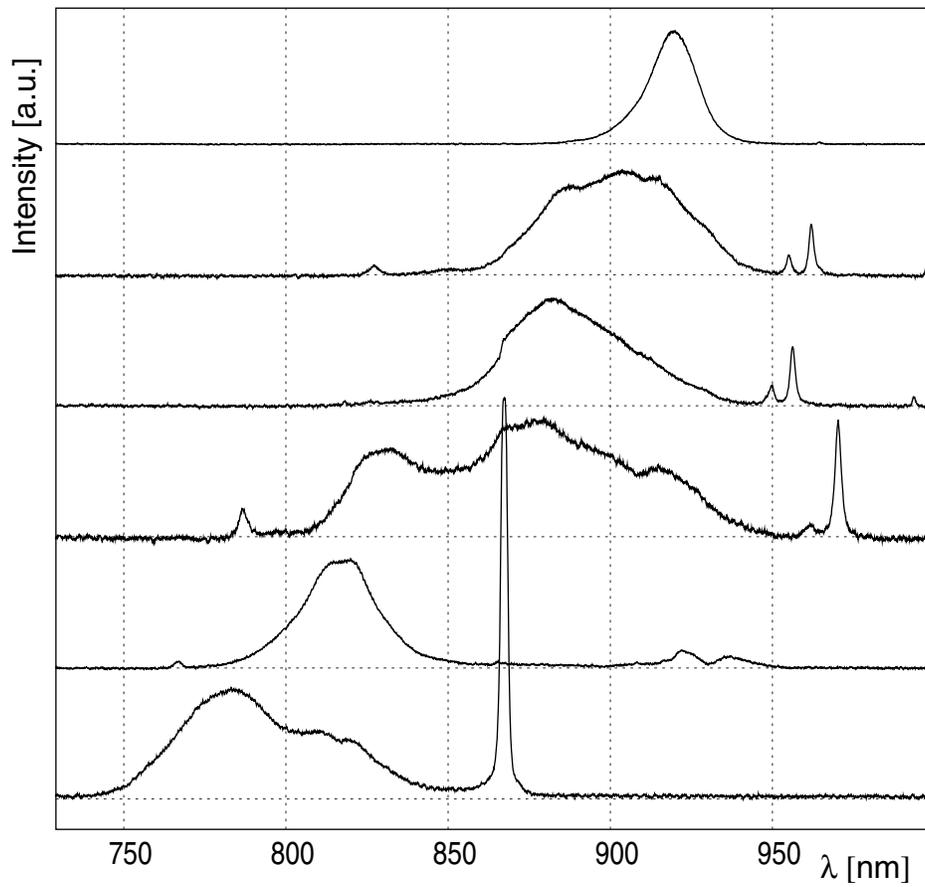

Figure 5: Spectra taken at different insertion depths of prism pair $P_{3/4}$ of the laser operating at the first stability limit. Modest CW breakthrough is sometimes seen (see text for further details).